\documentclass{article}
\usepackage[german, english]{babel}
\usepackage[a4paper, left=2.5cm, right=2.5cm, top=3.5cm, bottom=3cm]{geometry}
\usepackage{amssymb}
\usepackage{amsmath}
\usepackage{amscd}
\usepackage{latexsym}
\usepackage{graphicx}
\usepackage{ifthen}
\usepackage{epsfig}
\usepackage{setspace}
\usepackage{subcaption}
\usepackage{hyperref}
\hypersetup{
 colorlinks=true,
 linkcolor=blue,
 citecolor=magenta,
}

\catcode`@=11
\renewcommand{\section}{\@startsection{section}{1}{0pt}{\medskipamount}
{\medskipamount}{\large\bf}}
\numberwithin{equation}{section}
\catcode`@=12

\newcommand{\R}{\mathbb R}

\def\im{\mathrm{i}}
\def\ep{\mathrm{e}}

\def\diff{\mathrm{d}}

\def\sfrac#1#2{{\textstyle\frac{#1}{#2}}}
\def\>{\rangle}
\def\<{\langle}
\def\+{\dagger}
\def\={\ =\ }
\def\und{\quad\textrm{and}\quad}
\def\with{\quad\textrm{with}\quad}
\def\for{\quad\textrm{for}\quad}

\begin{document}

\title{\hbox{\bf A new construction of rational electromagnetic knots}}
\date{~}

\author{\phantom{.}\\[12pt]
{\Large Olaf Lechtenfeld}$^{\dagger}$ \
and \ {\Large Gleb Zhilin}$^{\dagger \star}$
\\[24pt]
$^{\dagger}${Institut f\"ur Theoretische Physik
and Riemann Center for Geometry and Physics}\\ 
{Leibniz Universit\"at Hannover} \\ 
{Appelstra{\ss}e 2, 30167 Hannover, Germany}
\\[12pt] 
$^{\star}${Department of Theoretical Physics and Astrophysics}\\
{Belarusian State University, Minsk 220004, Belarus}
} 

\clearpage
\maketitle
\thispagestyle{empty}

\begin{abstract}
\noindent\large
We set up a correspondence between solutions of the Yang--Mills equations 
on $\R \times S^3$ and in Minkowski spacetime via de Sitter space. 
Some known Abelian and non-Abelian exact solutions are rederived. 
For the Maxwell case we present a straightforward algorithm to generate 
an infinite number of explicit solutions, with fields and potentials in Minkowski
coordinates given by rational functions of increasing complexity.
We illustrate our method with a nontrivial example.
\end{abstract}

\newpage
\setcounter{page}{1} 

\section{ Conformal equivalence of dS$_4$ to ${\cal I}{\times}S^3$ and two copies of $\R_+^{1,3}$}

\noindent
The present work is motivated by the recent paper \cite{lechtenfeld-popov} co-authored by one of us,
where analytic solutions of the Yang--Mills equations on four-dimensional de Sitter space~dS$_4$
are constructed. It is well known that de Sitter space can be realized as the single-sheeted hyperboloid
\begin{equation}
-Z_0^2 + Z_1^2 + Z_2^2 + Z_3^2 + Z_4^2 \=\ell^2
\end{equation}
embedded in five-dimensional Minkowski space $\R^{1,4}$ with the metric
\begin{equation}
\mathrm{d}s^2 \= -\mathrm{d}Z_0^2 + \mathrm{d}Z_1^2 + \mathrm{d}Z_2^2 
+ \mathrm{d}Z_3^2 + \mathrm{d}Z_4^2\ .
\end{equation}
Constant $Z_0$ slices of the hyperboloid reveal a three-sphere of varying radius. 
The following parametrization makes this structure explicit:
\begin{equation}\label{r1s3}
Z_0 \= -\ell\,\cot\tau \quad\und\quad
Z_A \= \frac{\ell}{\sin\tau}\,\omega_A \for A = 1,\ldots,4\ ,
\end{equation}
where the coordinates $\omega_A$ embed a unit three-sphere into $R^4$, 
and $0<\tau<\pi$, i.e.
\begin{equation}
\omega_A \omega_A = 1 \quad\und\quad \tau\in{\cal I}:=(0,\pi)\ .
\end{equation}
The metric of dS$_4$ in such coordinates becomes
\begin{equation}\label{metriccyl}
\mathrm{d}s^2 \= \frac{\ell^2}{\sin^2\tau}
\bigl(-\mathrm{d}\tau^2 + \mathrm{d}\Omega_3^2\bigr)\ ,
\end{equation}
where $\diff\Omega_3^2$ denotes the metric of the unit three-sphere.
Hence, four-dimensional de Sitter space is conformally equivalent to
a finite Minkowskian cylinder over a three-sphere.

Part of it is also conformally equivalent to (half of) Minkowski space,
by employing the parametrization
\begin{equation}\label{r13}
Z_0 \= \frac{t^2-r^2-\ell^2}{2\,t}\ ,\quad 
Z_1 \= \ell\,\frac{x}{t}\ ,\quad 
Z_2 \= \ell\,\frac{y}{t}\ ,\quad 
Z_3 \= \ell\,\frac{z}{t}\ ,\quad 
Z_4 \= \frac{r^2-t^2-\ell^2}{2\,t}\ ,
\end{equation} 
where
\begin{equation}
x,y,z\in\R \quad\und\quad r^2 = x^2 + y^2 + z^2 \qquad\textrm{but}\qquad t\in\R_+
\end{equation}
since $t\to0$ corresponds to $Z_0\to-\infty$. The metric of dS$_4$ becomes
\begin{equation}\label{metricMink}
\mathrm{d}s^2 \=
\frac{\ell^2}{t^2}\,\bigl(-\mathrm{d}t^2 +\mathrm{d}x^2 +\mathrm{d}y^2 +\mathrm{d}z^2\bigr)\ ,
\end{equation}
hence these coordinates cover the future half $\R_+^{1,3}$ of Minkowski space. In a moment this parametrization will be extended to the whole of Minkowski space, 
by gluing a second copy of dS$_4$ to provide for the $t<0$ half.
The de Sitter radius~$\ell$ provides a scale.

We shall need the direct relation between the cylinder and Minkowski coordinates.
By comparing (\ref{r1s3}) and (\ref{r13}) we see that
\begin{equation}\label{cottau}
-\cot\tau \= \frac{t^2 - r^2 - \ell^2}{2\,\ell\,t}\ ,\quad
\omega_1 \= \gamma\,\frac{x}{\ell}\ ,\quad 
\omega_2 \= \gamma\,\frac{y}{\ell}\ ,\quad 
\omega_3 \= \gamma\,\frac{z}{\ell}\ ,\quad 
\omega_4 \= \gamma \frac{r^2 - t^2 - \ell^2}{2\,\ell^2}\ ,
\end{equation}
where for convenience we abbreviated the frequent combination
\begin{equation}\label{gamma}
\gamma \= \frac{2\,\ell^2}{\sqrt{4\,\ell^2 t^2 + (r^2-t^2+\ell^2)^2}}\ .
\end{equation}
If we fix $r$ and let $t$ vary from $-\infty$ to $\infty$, then $-\cot\tau$ sweeps two branches. 
We pick the branches so that $\tau\in(-\pi,0)$ for $t<0$ and $\tau\in(0,\pi)$ for $t>0$, gluing them
at $\tau=t=0$. Then inverting (\ref{cottau}) produces $\tau$ as a regular function of $(t,x,y,z)$. 
A more useful relation for the following is
\begin{equation}\label{expitau}
\exp (\mathrm{i}\,\tau) \= \frac{(\ell+\mathrm{i}t)^2+r^2}{\sqrt{4\,\ell^2 t^2 + (r^2-t^2+\ell^2)^2}}\ .
\end{equation} 
Hence, comparing (\ref{metriccyl}) and (\ref{metricMink}), 
we have given an explicit conformal equivalence between full Minkowski space $\R^{1,3}$
and a patch of a finite $S^3$-cylinder $2{\cal I}{\times}S^3$ with $2{\cal I}=(-\pi,\pi)\ni\tau$. The structure of this equivalence is best clarified by an illustration. Note that the whole infinite $R\times S^3$ cylinder can be covered by such patches. The neighbouring patches can be related via shifting $\tau$ by $\pi$ and changing the sign of $\omega_4$. The latter action essentially implements a parity transformation.  

\begin{figure}[h!]
\centering
\includegraphics[width = 0.6\paperwidth]{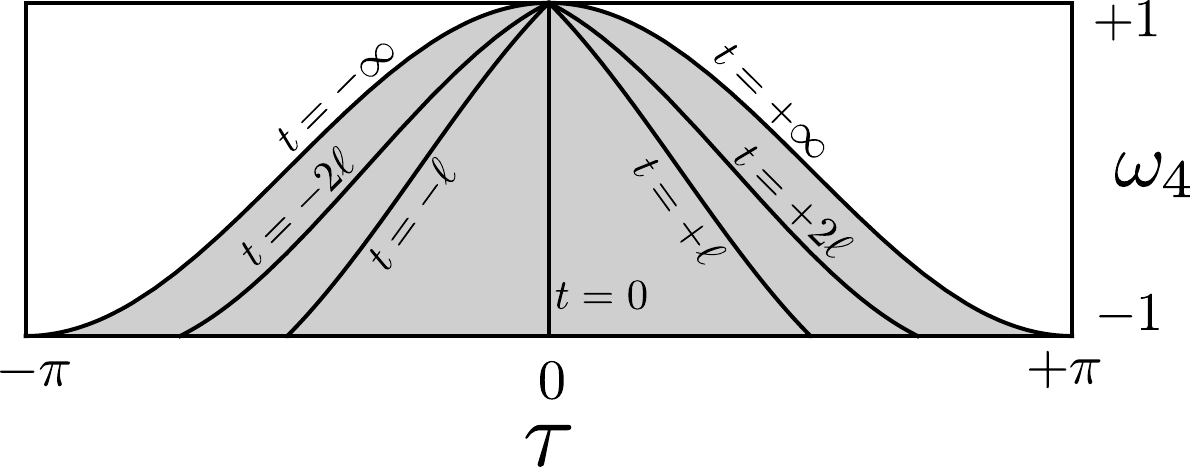}
\caption{An illustration of the map between a cylinder $2\mathcal{I}\times S^3$ and Minkowski space $R^{1,3}$.
The Minkowski coordinates cover the shaded area. The boundary of this area is given by the curve $\omega_4 = \cos\tau$. Each point is a two-sphere spanned by $\omega_{1,2,3}$, which is mapped to a sphere of constant $r$ and $t$.  }
\end{figure}

\section{The correspondence}

\noindent
In four spacetime dimensions Yang--Mills theory is conformally invariant. 
Therefore, instead of solving its equations of motion on Minkowski space
one may solve them on the cylinder $2{\cal I}{\times}S^3$. 
The latter has the added advantage yielding a manifestly SO(4)-covariant formalism
due to the three-sphere.
Furthermore, $S^3$ is the group manifold of~SU(2), 
which enables the geometric parametrization (we pick the temporal gauge $A_\tau=0$)
\begin{equation}\label{Aansatz}
A \= \sum\limits_{a=1}^{3} X_a(\tau,\omega) \, e^a\ ,
\end{equation}
where $X_a$ are three functions of $\tau$ and $\omega\equiv\{\omega_A\}$ 
valued in some Lie algebra, and $e^a$ are the three left-invariant one-forms on $S^3$. 
Since the conformal factor is irrelevant for the Yang--Mills equations
we can translate Yang--Mills solutions on $2{\cal I}{\times}S^3$ to solutions
on $\R^{1,3}$ simply via a change of coordinates. The behavior at the boundary $\cos\tau = \omega_4$ is thereby transferred to fall-off properties at temporal infinity $t\rightarrow \pm\infty$.

To become explicit, we need Minkowski-coordinate expressions 
for the one-forms~$e^0=\diff\tau$ and $e^a$, which are subject to
\begin{equation}
\mathrm{d}e^a + \varepsilon^a_{\ bc}\,e^b\wedge e^c \=0 \quad\und\quad
e^a e^a \= \diff\Omega_3^2\ .
\end{equation}
They can be constructed as
\begin{equation}
e^a \= - \eta^a_{BC} \ \omega_B\,\mathrm{d}\omega_C\ ,
\end{equation}
with $\eta^a_{BC}$ denoting the self-dual 't Hooft symbol (with non-zero components 
$\eta^i_{jk} = \varepsilon^i_{\ jk} \ \textrm{and}\ \eta^i_{j4} = -\eta^i_{4j} = \delta^i_j$). 
A straightforward computation yields ($a,j,k = 1,2,3$)
\begin{equation}\label{basismink}
\begin{aligned}
e^0 &\= \frac{\gamma^2}{\ell^3}\Bigl( 
\sfrac12(t^2 + r^2 + \ell^2)\,\mathrm{d}t - t\,x^k \mathrm{d}x^k \Bigr)\ ,\\
e^a &\= \frac{\gamma^2}{\ell^3}\Bigl( 
t\,x^a \mathrm{d}t - \bigl(\sfrac12(t^2 - r^2 + \ell^2)\,\delta^a_k + 
x^a x^k + \ell\,\varepsilon^a_{\ jk}x^j \bigr)\,\mathrm{d}x^k \Bigr)\ ,
\end{aligned}
\end{equation}
where we introduce the standard notation 
\begin{equation}
(x^i)=(x,y,z) \qquad\textrm{and (for later)}\qquad
(x^\mu)=(x^0,x^i)=(t,x,y,z)\ .
\end{equation}
Two remarks are in order. First, in Minkowski spacetime the parameter $\ell$ just sets an overall scale, 
which is needed for nontrivial solutions because the Yang--Mills equations themselves 
are scale-invariant in four dimensions. Second, at fixed $t$ the components for $e^0,\ldots,e^3$ 
decay at least as $1/r^2$ for large $r$. This is a good signal that the solutions translated from
the cylinder will have finite energy in $\R^{1,3}$.

Let us see how this works by transferring some solutions obtained in 
\cite{lechtenfeld-popov, lechtenfeld-popov2} to Minkowski spacetime.\footnote{
In these papers cylinder solutions were transferred to dS$_4$ solutions.}
There, the authors restricted to SO(4)-symmetric configurations by
taking $X_a=X_a(\tau)$ to be independent of~$\omega$.
This ansatz reduces the Yang--Mills equations to ordinary differential equations 
for the matrices $X_a$. 
On the cylinder, a simple static homogeneous solution is given by
\begin{equation}\label{saddle}
X_a (\tau) \= \sfrac{1}{2}\,T_a \qquad\Rightarrow\qquad
A \= \sfrac12\,g^{-1}\diff g \quad\for g:S^3\to\textrm{SU}(2)\ ,
\end{equation}
where $T_a$ are $su(2)$ algebra generators scaled to obey $[T_a,T_b] = 2\varepsilon_{abc} T_c$. 
After inserting (\ref{basismink}) and (\ref{saddle}) into the ansatz (\ref{Aansatz}) one recognizes 
the De Alfaro--Fubini--Furlan solution~\cite{dealfaro} (see also \cite{gibbons}). A more general case,
\begin{equation}
X_a(\tau) \= \bigl(1 + \sfrac12q(\tau)\bigr)\,T_a \qquad\textrm{with}\qquad
\frac{\mathrm{d}^2q}{\mathrm{d}\tau^2}\=-\frac{\partial V}{\partial q} 
\for V(q)\=\sfrac12q^2(q{+}2)^2\ ,
\end{equation} 
produces a family of SO(4)-symmetric solutions studied by L\"uscher~\cite{luscher}. 
For a review on analytic Yang--Mills solutions, see~\cite{actor}.

However, the interest of this paper is in Abelian solutions, i.e. electromagnetic field configurations.
These may be embedded in the non-Abelian framework by demanding the three matrices $X_a$ 
to all be proportional to the same fixed Lie-algebra element, say~$T_3$. 
Such solutions on ${\cal I}{\times}S^3$ (with two proportionality coefficients vanishing) 
were also discussed in~\cite{lechtenfeld-popov2}.
Since in the U(1)~case the matrix structure is irrelevant, from now on we take $X_a(\tau,\omega)$
simply to be real-valued functions and focus on Maxwell's equations.
In the SO(4)-invariant case, $X_a=X_a(\tau)$ are found to obey the oscillator equation
\begin{equation}
\frac{\mathrm{d}^2}{\mathrm{d}\tau^2}X_a(\tau) \= -4\,X_a(\tau)
\qquad\Rightarrow\qquad
X_a(\tau) \= c_a\,\cos\bigl(2(\tau{-}\tau_a)\bigr)\ ,
\end{equation} 
yielding six integration constants in the general solution.
Since the $X_a$ are oscillating with a frequency of two, we can use the simple expression 
(\ref{expitau}) for $\mathrm{e}^{2\mathrm{i}\tau}$ to translate the dependence on $\tau$ into
a rational expression in $t$ and~$r$. 
From
\begin{equation}
A \= X_a(\tau)\,e^a \= A_\mu\,\diff x^\mu
\end{equation}
one gets the components $A_\mu$ of the gauge potential 
after substituting the expressions (\ref{basismink}) for $e^a$.
Note that we chose $A_\tau=0$ but $A_t=A_0$ will be nonvanishing.
Likewise, from
\begin{equation}
\mathrm{d}A \= \frac{\diff}{\diff\tau}X_a(\tau)\,e^0\wedge e^a 
-\varepsilon^a_{\ bc} X_a(\tau)\,e^b \wedge e^c 
\= \sfrac{1}{2} F_{\mu \nu}\,\diff x^\mu \wedge \diff x^\nu 
\end{equation}
one can extract the electric and magnetic field components as 
\begin{equation}
E_i \= F_{i0} \quad\und\quad B_i \= \sfrac{1}{2}\varepsilon_{ijk}F_{jk}\ .
\end{equation}

As a particular example, let us pick the solution
\begin{equation}
X_1(\tau) \= -\sfrac{1}{8} \sin 2\tau\ ,\quad 
X_2(\tau) \= -\sfrac{1}{8} \cos 2\tau\ ,\quad 
X_3(\tau) \= 0
\end{equation}
and put $\ell=1$ for simplicity. 
A short computation leads to the electromagnetic field
\begin{equation} \label{knotEB}
\vec{E} + \mathrm{i} \vec{B} \= \frac{1}{\bigl((t-\mathrm{i})^2-r^2\bigr)^3}
\begin{pmatrix}
(x-\mathrm{i}y)^2-(t-\mathrm{i}-z)^2 \\
\mathrm{i}(x-\mathrm{i}y)^2 + \mathrm{i}(t-\mathrm{i}-z)^2 \\
-2\,(x-\mathrm{i}y)\,(t-\mathrm{i}-z)
\end{pmatrix}\ .
\end{equation}
This is the celebrated Hopf-Ra\~nada electromagnetic knot~\cite{elknots}.  
Note that in our approach we can also reconstruct the gauge potential for this solution.

\section{Construction of electromagnetic solutions}

\noindent
We will now investigate the case of a U(1) gauge group more thoroughly,
by allowing for a general dependence on the Minkowski coordinates~$x^\mu$ 
beyond one dictated by SO(4)~invariance like in (\ref{knotEB}) above.
This means that our functions~$X_a$ will also depend on the $S^3$ coordinates,
$X_a=X_a(\tau,\omega)$.
In order to capture this dependence in an SO(4)-covariant fashion, 
let us introduce a basis of three-sphere vector fields dual to the left-invariant one-forms~$e^a$,
\begin{equation} \label{V^R}
R_a \= - \eta^a_{BC}\,\omega_B \frac{\partial}{\partial \omega_C}\ .
\end{equation}
Under commutation, these vectors fields form an $su(2)$ representation,
\begin{equation}
[R_a,R_b] \= 2\,\varepsilon_{abc}\,R_c\ .
\end{equation}
They realize the infinitesimal right multiplications on SU(2), hence the chosen notation. 
Hence for an arbitrary function $\Phi$ on $S^3$ we may write
\begin{equation}
\mathrm{d}\Phi \= e^a\,R_a \Phi\ .
\end{equation}
There is another triplet of vector fields closely related to (\ref{V^R}) given by
\begin{equation}
L_a = - \tilde{\eta}^a_{BC}\,\omega_B \frac{\partial}{\partial \omega_C}\ ,
\end{equation}
where $\tilde{\eta}^a_{BC}$ is the anti-self-dual 't Hooft symbol 
(with non-zero components $\eta^i_{jk} = \varepsilon^i_{\ jk}$ and $\eta^i_{j4} = -\eta^i_{4j} = -\delta^i_j$). 
As suggested by the notation, these realize infinitesimal left multiplications 
and obey the same algebra as the $R$'s. Since right and left multiplications commute, 
$[R_a,L_b] = 0$.

The space of functions of $S^3$ can be decomposed into irreducible representations of 
the $su(2)_L\times su(2)_R$ algebra generated by $L_a$ and $R_b$, respectively. 
These representations can be labeled uniquely by a non-negative number $j$
such that $2j\in\{0,1,2,\ldots\}$.
If we define Hermitian ``angular momentum'' operators 
\begin{equation}
I_a\ :=\ \sfrac{\im}{2}\,L_a \quad\und\quad J_a\ :=\ \sfrac{\im}{2}\,R_a\ ,
\end{equation}
then a particular basis of hyperspherical harmonics 
\begin{equation}
Y_{j; m,n}(\omega) \qquad\with\quad m,n = -j,-j{+}1,\ldots,+j \und 2j=0,1,2,\ldots
\end{equation}
is specified by the relations
\begin{equation}
\begin{aligned}
I_3\,Y_{j; m,n} &\= m\,Y_{j; m,n}\ ,\qquad J_3\,Y_{j; m,n} \= n\,Y_{j; m,n}\ ,\\[4pt]
I^2\,Y_{j; m,n} &\= J^2\,Y_{j; m,n} \= j(j{+}1)\,Y_{j; m,n}\ ,
\end{aligned}
\end{equation}
where $I^2=I_a I_a$ and $J^2=J_a J_a$ are the Casimirs of the two $su(2)$ subalgebras.
To give explicit formul\ae\ for $Y_{j; m,n}$ it is useful to introduce two complex coordinates
\begin{equation}
\alpha = \omega_1 + \mathrm{i} \omega_2 \qquad\und\qquad
\beta = \omega_3 + \mathrm{i} \omega_4
\end{equation}
parametrizing the three-sphere via $\bar{\alpha}\alpha+\bar{\beta}\beta=1$.
Then, employing the notation
\begin{equation}
I_\pm = (I_1\pm\im I_2)/\sqrt{2}\ ,\quad J_\pm = (J_1\pm\im J_2)/\sqrt{2} 
\quad\und\quad X_\pm = (X_1\pm\im X_2)/\sqrt{2}\ ,
\end{equation}
the angular momentum generators take the simple form
\begin{align}
I_+ &\= (\bar{\beta} \partial_{\bar{\alpha}} - \alpha\partial_\beta)/\sqrt{2}\ ,
&J_+& \= (\beta\partial_{\bar{\alpha}} - \alpha\partial_{\bar{\beta}})/\sqrt{2}\ , \nonumber \\
I_3 \,&\= (\alpha \partial_\alpha + \bar{\beta} \partial_{\bar{\beta}} - \bar{\alpha}\partial_{\bar{\alpha}} - \beta\partial_{\beta})/2\ ,
&J_3\,& \= (\alpha \partial_\alpha + \beta\partial_{\beta}- \bar{\alpha}\partial_{\bar{\alpha}} - \bar{\beta} \partial_{\bar{\beta}} )/2\ ,  \\
I_- &\= (\bar{\alpha}\partial_{\bar{\beta}} - \beta\partial_\alpha)/\sqrt{2}\ ,
&J_-& \= (\bar{\alpha}\partial_\beta - \bar{\beta}\partial_\alpha)/\sqrt{2} \ ,\nonumber
\end{align}
and the normalized hyperspherical harmonics are
\begin{equation}
Y_{j;m,n} \= 
\sqrt{\frac{2j{+}1}{2 \pi^2}} \sqrt{\frac{2^{j{-}m}(j{+}m)!}{(2j)!\,(j{-}m)!} \frac{2^{j{-}n}(j{+}n)!}{(2j)!\,(j{-}n)!}}\, 
(I_{-})^{j-m} (J_{-})^{j-n}\,\alpha^{2j}\ .
\end{equation}
These are homogeneous polynomials of degree~$2j$ 
in $\alpha$ and $\beta$ and their complex conjugates.

We will work in the Coulomb gauge on $2{\cal I}\times S^3$, which for a general gauge field
\begin{equation}
A \= X_0(\tau,\omega)\,\mathrm{d}\tau + X_a(\tau,\omega)\,e^a
\end{equation}
with $\tau\in(-\pi,+\pi)$ and $\omega\in S^3$ means that
\begin{equation}
X_0(\tau,\omega) = 0 \quad\und\quad J_a\,X_a (\tau,\omega) = 0\ .
\end{equation}
The Maxwell equations of motion then take the form
\begin{equation}\label{MaxwellX}
-\sfrac14\,\partial_\tau^2 X_a \= (J^2{+}1)\,X_a + 2\im\,\varepsilon_{abc} J_b\,X_c\ ,
\end{equation}
which is $su(2)_L$ invariant and $su(2)_R$ covariant. 
A less compact but more transparent rewriting of Maxwell's and our gauge-fixing equation is
\begin{align}
-\sfrac14\,\partial_\tau^2 X_+ &\=  (J^2-J_3+1)\,X_+ + J_+ X_3 \ , \nonumber \\[4pt]
-\sfrac14\,\partial_\tau^2 X_3\, &\=  (J^2+1)\,X_3 - J_+ X_- + J_- X_+ \ , \label{Maxwell}\\[4pt]
-\sfrac14\,\partial_\tau^2 X_- &\=  (J^2+J_3+1)\,X_- - J_- X_3  \ , \nonumber \\[4pt]
0 &\= J_3 X_3 + J_+ X_- + J_- X_+\ .\label{gaugefix}
\end{align}
The components $X_a$ are functions on $S^3$ and can be expanded in the basis of hyperspherical harmonics,
\begin{equation} \label{generalX}
X_a (\tau,\omega) \= \sum_{jmn} X_a^{j;m,n}(\tau)\,Y_{j;m,n}(\alpha,\beta)\ .
\end{equation}
From the form of (\ref{Maxwell}) and~(\ref{gaugefix}) it is obvious that 
\begin{itemize}
\addtolength{\itemsep}{-4pt}
\item the equations are diagonal in the quantum numbers $j$ and $m$, so these may be kept fixed
\item they only couple triplets $(X_3^{j;m,n}, X_+^{j;m,n{+}1}, X_-^{j;m,n-1})$, 
so $X_\pm\propto J_\pm X_3$ for $X_3\propto Y_{j;m,n}$
\item the ansatz $X_a^{j;m,n}(\tau)\propto\ep^{\im\Omega_a^{j;n}\tau}c_a^{j;n}$ 
gives a linear system for the frequencies and amplitudes
\end{itemize}

It turns out that the former are integral,
\begin{equation}
\Omega_a^{j;n} \= \pm2(j{+}1) \quad\textrm{or}\quad \pm2j
\end{equation}
independent of $a$ or~$n$. We call the corresponding solutions `type I' and `type II', respectively.
From the linear equations we extract the coefficients $c_a^{j;n}$ 
(up to an irrelevant overall factor which may depend on $j$ and~$m$).
Putting everything together, the two families of basic solutions read:
\begin{itemize}
\item type I : \quad 
$j \geq 0\ ,\quad m = -j,\ldots,+j\ ,\quad n = -j{-}1,\ldots,+j{+}1\ ,\quad \Omega^j=\pm2(j{+}1)\ ,$
\begin{equation}
\begin{aligned}
X_+  &\= \sqrt{(j{-}n)(j{-}n{+}1)/2} \ \mathrm{e}^{\pm2(j+1)\mathrm{i}\tau} \ Y_{j;m,n+1}\ , \\
X_3\,&\= \sqrt{(j{+}1)^2-n^2} \ \mathrm{e}^{\pm2(j+1)\mathrm{i}\tau} \ Y_{j;m,n}\ , \\
X_-   &\= -\sqrt{(j{+}n)(j{+}n{+}1)/2} \ \mathrm{e}^{\pm2(j+1)\mathrm{i}\tau} \ Y_{j;m,n-1}\ .
\end{aligned}
\end{equation}
\item type II :\quad
$j \geq 1\ ,\quad m = -j,\ldots,+j\ ,\quad n = -j{+}1,\ldots,+j{-}1\ ,\quad \Omega^j=\pm2j\ ,$
\begin{equation}
\begin{aligned}
X_+  &\= -\sqrt{(j{+}n)(j{+}n{+}1)/2} \ \mathrm{e}^{\pm2j\,\mathrm{i}\tau} \ Y_{j;m,n+1}\ , \\
X_3\,&\= \sqrt{j^2-n^2} \ \mathrm{e}^{\pm2j\,\mathrm{i}\tau} \ Y_{j;m,n}\ , \\
X_-   &\= \sqrt{(j{-}n)(j{-}n{+}1)/2} \ \mathrm{e}^{\pm2j\,\mathrm{i}\tau} \ Y_{j;m,n-1}\ .
\end{aligned}
\end{equation}
\end{itemize}
It is understood that $Y_{j;m,n}$ vanish when $n$ lies outside the interval $[-j,+j]$.
This occurs on type~I, at the maximal $n$ values (two $X$ components vanish) and at the
next-to-maximal $n$ values (one $X$ component vanishes).
We see that, for a given integral frequency $\Omega$, solutions occur with the two values
$j=\sfrac12|\Omega|$ and $j=\sfrac12|\Omega|{-}1$ (except for $\Omega{=}1$). 
Constant solutions ($\Omega{=}0$) are not admissible. 
The simplest nontrivial case are the six SO(4)-invariant $j{=}0$ solutions of type~I.
Indeed,  a superposition of $(j;m,n)=(0;0,-1)$ and $(0;0,+1)$
was presented at the end of the previous section.

\section{ Some properties of the solutions}

\noindent
The basis solutions given in the previous section are complex, 
but real and imaginary parts solve the equations independently. 
For fixed $j$ there are $2(2j{+}1)(2j{+}3)$ real linearly independent solutions of type~I 
and $2(2j{+}1)(2j{-}1)$ such solutions of type II (with $j{>}0$). 
In total, for $j>0$ there are $4 (2j{+}1)^2$ solutions, and for $j=0$ there are six,
in agreement with SO(4) representation theory.
Type~I solutions at level $j$ are related with type~II solutions at level $j{+}1$ 
via a parity transformation. On $S^3$ this transformation exchanges right and left $su(2)$ algebras,
which induces $n\leftrightarrow m$ for the hyperspherical harmonics. 
The shift in $j$ arises from expanding the right-invariant one-forms in terms of the left-invariant ones.
Furthermore, electromagnetic duality is realized in a simple fashion for the solutions above. 
For type I (type II) at fixed $j$, shifting $|\Omega^j|\tau$ by $\frac{\pi}{2}$ ($-\frac{\pi}{2}$) 
produces the dual solution with $\vec{B}_D = \vec{E}$ and $\vec{E}_D = - \vec{B}$.
All quantum numbers remain unaffected.

The field energy and helicity, given by
\begin{equation} \label{enhel}
E \= \sfrac12 \int \limits_{\R^3} \!\diff^3\! x \ \bigl(\vec{E}^2 + \vec{B}^2\bigr) 
\quad\und\quad
h \= \sfrac12 \int \limits_{\R^3} \ \bigl( A\wedge F + A_D \wedge F_D \bigr)\ ,
\end{equation}
respectively, are both conserved in the dynamics.
They are readily computed for any given solution.
If we introduce ``sphere frame'' electric and magnetic fields (denoted calligraphically) via
\begin{equation}
F \=  \mathcal{E}_a\,e^a \wedge e^0 + \sfrac{1}{2} \mathcal{B}_a\,\varepsilon^a_{\ bc}\,e^b\wedge e^c\ ,
\end{equation}
then on the $t = \tau = 0$ slice we find that
\begin{equation}
\int \limits_{\R^3} \!\diff^3\! x \ \vec{E}^2 \= 
\frac{1}{\ell}\int \limits_{S^3} \!\diff^3\Omega_3 \  (1{-}\omega_4)\,\mathcal{E}_a\mathcal{E}_a 
\quad\und\quad
\int \limits_{\R^3} \!\diff^3 \!x \ \vec{B}^2 \= 
\frac{1}{\ell}\int \limits_{S^3} \!\diff^3\Omega_3 \  (1{-}\omega_4)\,\mathcal{B}_a \mathcal{B}_a \ .
\end{equation}
Manipulating integrals on the sphere is more convenient due to the orthogonality properties 
of the harmonics $Y_{j;m,n}$. 
For a solution with fixed $(j,m,n)$ the $\omega_4$ term drops out from the integration. 
In the expression (\ref{enhel}) for the helicity the metric does not enter, so we may evaluate
the corresponding integral on any spatial slice of the cylinder $2{\cal I}\times S^3$.
The dual potential is easily accessible, as was remarked in the previous paragraph.
Energy and helicity are related: for instance, any type~I solution with $m=n=0$ and fixed~$j$ has
$E\,\ell=2(j{+}1)\,h$.

The main technical but straightforward task is to transform any chosen solution (\ref{generalX})
to Minkowski coordinates $(t,x,y,z)$. 
Since the hyperspherical harmonics are homogeneous polynomials in $\omega_A$, 
one can easily use the expressions (\ref{cottau}) and~(\ref{gamma}) to do this.
The remaining coordinate dependence hides in integral powers of $\mathrm{e}^{\mathrm{i}\tau}$,
for which we employ formula~(\ref{expitau}).
All resulting expressions will be rational functions in the Minkowski coordinates.
Potential square roots stemming from odd powers of $\mathrm{e}^{\mathrm{i}\tau}$ or $\gamma$
will not occur in the final expressions for the components $A_\mu(x,y,z,t)$, as one can check.

\section{An example}

\noindent
Let us illustrate the power of our method by presenting a simple example of a solution 
with a nontrivial dependence on the $S^3$~coordinates. 
We take the two $j=1$ solutions of type~I with $m=n=0$ and combine their $\ep^{\pm4\im\tau}$
time dependences into a cosine:
\begin{equation}
X_+ \= -\sfrac{\sqrt{3}}{\pi}\,\alpha\beta\,\cos 4\tau \ ,\qquad 
X_3 \= \sfrac{\sqrt{6}}{\pi}\,(\beta\bar{\beta}-\alpha\bar{\alpha})\,\cos 4\tau \ ,\qquad 
X_- \= -\sfrac{\sqrt{3}}{\pi}\,\bar{\alpha}\bar{\beta}\,\cos 4\tau \ .
\end{equation}
On the $t{=}\tau{=}0$ slice time and $\tau$ derivatives are proportional to one another, 
hence at $t{=}0$ this solution is purely magnetic.
We refrain from presenting here the rather lengthy expressions for the electromagnetic fields,
but instead display below plots of the energy density level surfaces and a magnetic flux line,
all at $t{=}0$. In this example, the energy and helicity compute to
\begin{equation}
E \= 48/\ell \quad\und\quad h \= 12\ .
\end{equation}
Finally, in the figure below we have added the energy density plot for a particular 
$(j;m,n)=(\sfrac32;\sfrac12,\sfrac32)$ type~I solution, for illustrative and aesthetical reasons.
\begin{figure}[h!]
\centering
\begin{subfigure}[b]{0.4\linewidth}
\includegraphics[width=\linewidth]{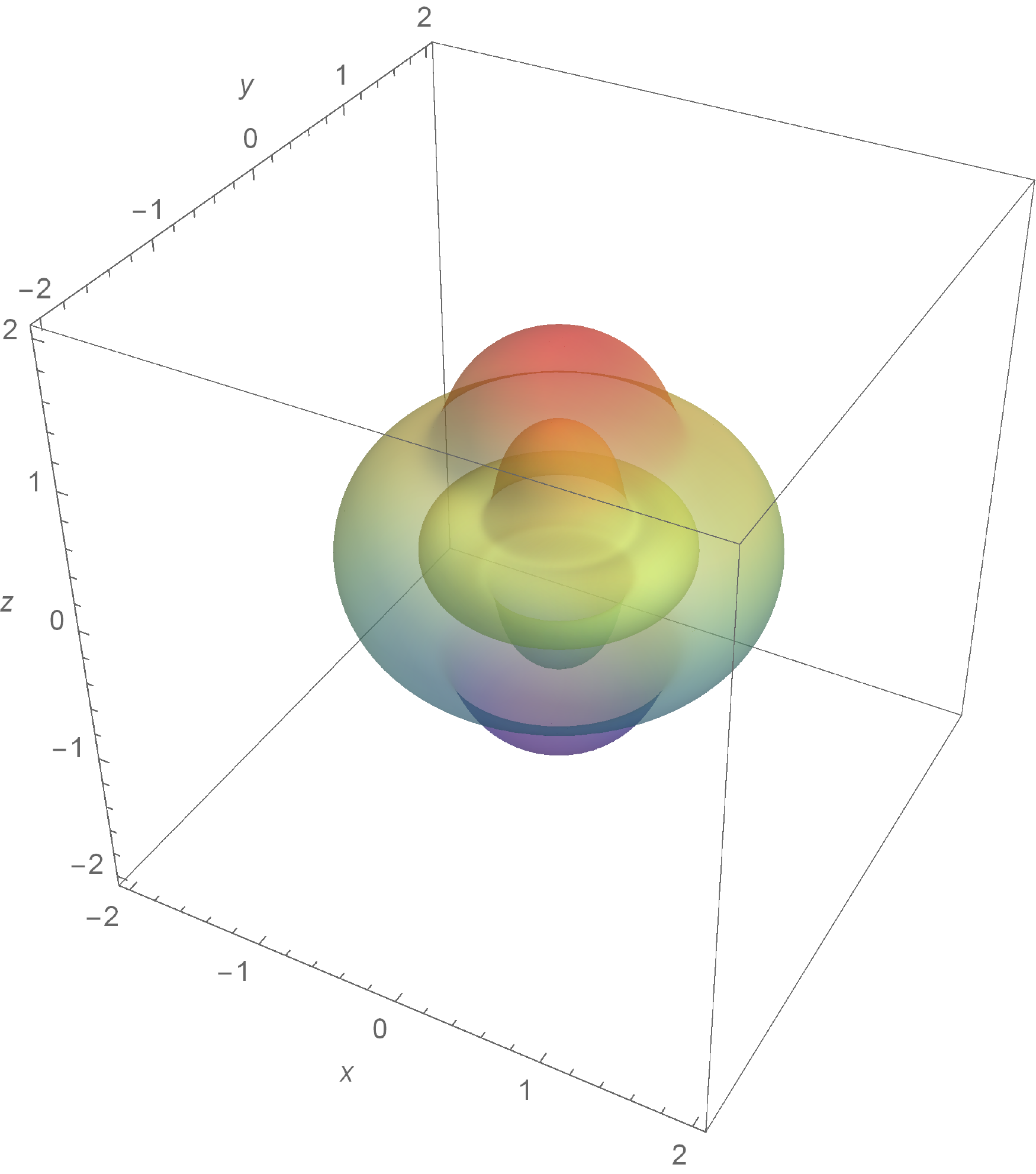}
\caption{}
\end{subfigure}
\begin{subfigure}[b]{0.4\linewidth}
\includegraphics[width=\linewidth]{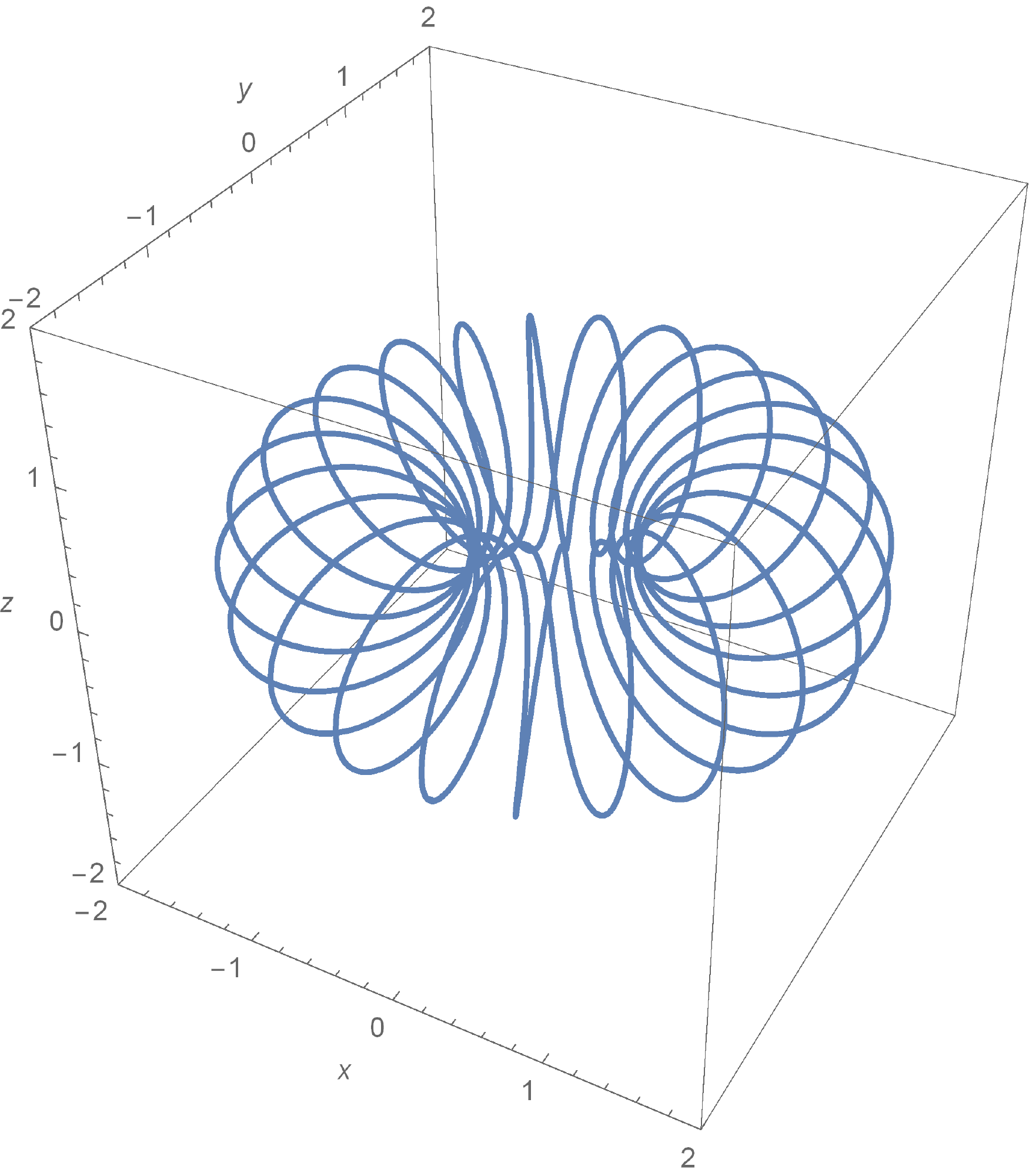}
\caption{}
\end{subfigure}
\begin{subfigure}[b]{0.4\linewidth}
\includegraphics[width=\linewidth]{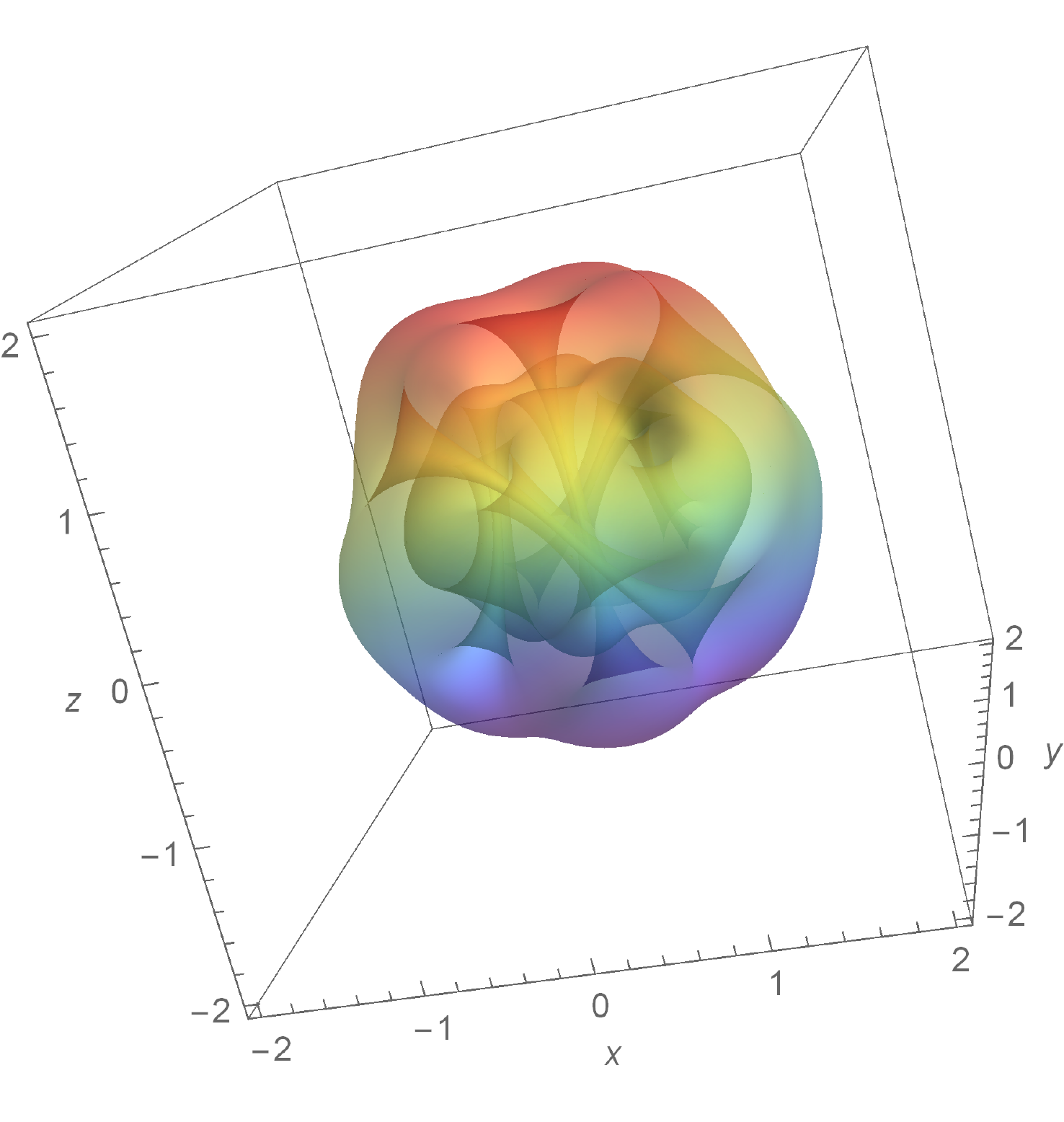}
\caption{}
\end{subfigure}
\caption{(a) Energy density level surfaces and (b) a particular closed magnetic field line 
for the solution discussed in the example. 
Plot (c) shows a particular $(\sfrac32;\sfrac12,\sfrac32)$ solution. 
The energy density levels displayed have $0.01$ and $0.1$ of the maximal value. 
All plots are at $t{=}0$.}
\end{figure}

\section{Summary and discussion}

\noindent
The construction of rational electromagnetic field configurations with nontrivial topology
has been an active field of theoretical and experimental research for almost thirty years now.
To the four methods described in the recent review~\cite{knotreview} we should like to add
a fifth one. It is based on the simplicity of analytically solving Maxwell's equations on
a temporal cylinder over a three-sphere plus the conformal equivalence of a part of the latter space
to four-dimensional Minkowski spacetime. 

A generic Maxwell solution on $\R\times S^3$, written in the form 
$A = X_\alpha(\tau,\omega)\,e^\alpha$ ($\alpha=0,\ldots,3)$, 
can be translated  via a change of coordinates to a Minkowski solution 
$A = X_\alpha\bigl(\tau(x),\omega(x)\bigr)\,{e^\alpha}_\mu(x)\,\mathrm{d}x^\mu$. 
We have demonstrated the effectiveness of this approach by reconstructing 
several known exact solutions in Minkowski spacetime.
The pre-image of full Minkowski spacetime lies in a finite open $(-\pi,+\pi)$ segment of
the cylinder $\R\times S^3$, so only finite-time dynamics is required there.

Suppose we are given an electromagnetic field in Minkowski spacetime at $t=0$ that decays 
quickly enough at infinity. It can be mapped to the $\tau{=}0$ slice of $\R\times S^3$ and 
consequently expanded in the complete basis there. After going back to Minkowski space 
we obtain a solution for all values of~$t$. In this sense the set of finite-energy solutions 
of the Maxwell equations that we have constructed forms a complete basis: 
any solution which at $t{=}0$ decays quickly enough at spatial infinity can be expanded in it.
Our simple and explicit expressions allow for the transfer of this complete basis on $S^3$ 
to a complete set of Maxwell configurations with sufficiently fast spatial and temporal decay 
on $\R^{1,3}$. By construction, these fields have finite energy and action. 

Our method works likewise for Yang--Mills fields. In fact, it was conceived first for non-Abelian 
gauge theory, and we have rediscovered exact Minkowskian SU(2) Yang--Mills solutions with it.
However, in the Yang--Mills equations of motion, the commutator term couples different $j$
components of $X_a$ in (\ref{generalX}), and so the analysis of an infinite coupled set of
now nonlinear ordinary differential equations generalizing~(\ref{MaxwellX}) will be much harder.
However, because of the simple and regular form of $\tau(x)$, $\omega(x)$ and ${e^\alpha}_\mu(x)$, 
one may hope that our method can become a useful tool for analytic and numerical investigations 
of classical Yang--Mills dynamics in four-dimensional Minkowski spacetime.

\section*{Acknowledgements}
\noindent
We are grateful to A.D.~Popov for helpful comments and pointing out literature, 
and to Y.M.~Shnir for useful discussions.
Z.G.~is grateful to Weizmann Institute of Science for hospitality. 
This work was partially supported by the Deutsche Forschungsgemeinschaft grant LE~838/13
and by the Research Training Group RTG 1463. This article is based upon work from COST
Action MP1405 QSPACE, supported by COST (European Cooperation in Science and Technology).

\end{document}